\documentstyle[aps,prb,epsfig,calc]{revtex}

\begin{document}
\title{Master equation approach to configurational
kinetics of non-equilibrium alloys and its application to studies of L1$_0$-type
orderings}

\author{K. D. Belashchenko\cite{Ames}, I. R. Pankratov, G. D. Samolyuk\cite{Ames} and V. G. Vaks} 

\address{Russian Research Centre ``Kurchatov Institute'', Moscow 123182, Russia}

\maketitle

\begin{abstract}

We review a series of works where the fundamental master equation is used
to develop a microscopical description of evolution 
of non-equilibrium atomic distributions in alloys. We describe exact 
equations for temporal evolution of local concentrations and their correlators
as well as approximate methods to treat these equations,
such as the kinetic mean-field and the kinetic cluster methods. We also describe 
an application of these methods to studies of kinetics of
L1$_0$-type orderings in FCC alloys which 
reveal a number of peculiar microstructural effects, many of them
agreeing well with experimental observations. 

\end{abstract}

\section{Introduction}\label{int}

Problems of evolution of non-equilibrium statistical systems
attract attention in many areas of physics. These problems are of particular interest for
configurational alloy  kinetics---the evolution of the atomic distribution 
in non-equilibrium alloys. The microstructure and macroscopic
properties of such alloys, e.g.  strength and plasticity, depend crucially on
their thermal and mechanical history---for example, on
the kinetic path taken during phase transformations.
Theoretical treatments of these problems usually  employ either 
Monte Carlo simulation---see  e.g. \cite{FVCP}---or
phenomenological kinetic equations for 
local concentrations and order parameters \cite{Khach-book,CWK-92,Khach-00}.
However, Monte Carlo studies in this field
are time consuming, and until now they provided a limited 
information on the details of the microstructural evolution.
Use of the phenomenological kinetic equations is more feasible, and 
Khachaturyan and co-workers \cite{Khach-book,CWK-92,Khach-00} used this approach 
as a basis for discussing many microstructural effects. However,
the phenomenological approach employs a number of unclear
approximations---in particular, the extrapolation of
linear Onsager equations for weakly nonequilibrium states  to
the nonlinear region of states far from equilibrium,
and the relation between the phenomenological and microscopic approaches 
is often unclear.

A consistent description of non-equilibrium alloys can be based 
on the fundamental master equation for the probabilities of various atomic 
distributions over lattice sites. The idea to employ this equation for
studies of phase transformations was first suggested by Martin
\cite{Martin}. For the last decade this approach has been formulated 
in terms of both approximate and exact kinetic equations 
\cite{V-Beiden,VBD,Vaks-96,BV-98,BDPSV,BDSV,Nastar}
and was applied to many concrete problems
\cite{V-Beiden,VBD,Vaks-96,BV-98,BDPSV,BDSV,Nastar,DMSV,DVM,%
BV-96,BV-97,DV,BSV,Vaks-01,PV}.
In this paper we describe the main ideas and methods of this approach
and illustrate them with an application
to studies of microstructural evolution  under L1$_0$ (CuAu I)-type  orderings
in FCC alloys.

\section{Exact relations}\label{exact-eqs}

Following Ref. \onlinecite{BV-98} we consider a substitutional alloy that
includes atoms of $m$ species
${\rm p}={\rm p}_1,{\rm p}_2,\ldots {\rm p}_m$,
in particular, vacancies for which p=v. Various distributions of 
atoms over lattice sites $i$ are described by the different
occupation number sets $\{n_{{\rm p}i}\}$ where the operator $n_{{\rm p}i}$
is unity when the site $i$ is occupied by a p-species atom and
 zero otherwise.  For each $i$ the sum of operators $n_{{\rm p}i}$ over all species p
is unity, thus only $m-1$ of them are independent.
It is convenient to mark  the independent operators with special symbols,
e.g. with greek letters: $(n_{{\rm p}i})_{\rm indep}=n_{\alpha i}$,  while
the rest operator  denoted as $n_{{\rm r}i}$ is expressed via
$n_{\alpha i}$:
\begin{equation}
n_{{\rm r}i}=1-\sum_{\alpha}n_{\alpha i}.
\label{n_r}
\end{equation}
The configurationally dependent part of energy $H^t$ can be  written as
\begin{equation}
H^t=\sum_{{\rm pq},i>j}V_{ij}^{{\rm p}{\rm q}}n_{{\rm p}i}n_{{\rm q}j}+
\sum_{{\rm pqr},i>j>k}V_{ijk}^{{\rm p}{\rm q}{\rm r}}n_{{\rm p}i}n_{{\rm q}j}n_{{\rm r}k}+
\ldots
\label{H^t}
\end{equation}
 After  elimination of the operators 
$n_{{\rm r}i}$ according to (\ref{n_r})
Eq. (\ref{H^t}) yields %the expression for
the interaction Hamiltonian $H$ in terms of only
independent operators $n_{\alpha i}$:
\begin{equation}
H=\sum_{\alpha\beta, i>j}v_{ij}^{\alpha\beta}n_{\alpha i}n_{\beta j}+
+\sum_{\alpha\beta\gamma, i>j>k}v_{ijk}^{\alpha\beta\gamma}n_{\alpha i}n_{\beta j}
n_{\gamma k}+
\ldots
\label{H}
\end{equation}
where effective interactions
$v_{i\ldots j}^{\alpha\ldots \beta}$ are linearly expressed via
$V_{i\ldots j}^{\rm p\ldots q}$ in (\ref{H^t}). 
% for example:
%\begin{equation}
%$v_{ij}^{\alpha\beta}=(V^{\alpha\beta}-V^{\alpha{\rm r}}-V^{{\rm r}\beta}
%+V^{\rm rr})_{ij}+\ldots $
%\label{v_ij}
%\end{equation}
%\end{numparts}

The fundamental master equation for the probability $P$
to find the occupation number set $\{n_{\alpha i}\}=\xi$ is
\begin{equation}
%{d\over dt}
dP(\xi)/dt=\sum_{\eta}
[W(\xi,\eta)P(\eta)-
W(\eta,\xi)P(\xi)]
\label{m-eq}
\end{equation}
where $W(\xi,\eta)$ is the $\eta\rightarrow\xi$ transition probability
per unit time.
Adopting  for this probabilitiy the conventional ``thermally activated
atomic exchange model'' \cite{Martin,V-Beiden},
we can express $W(\xi,\eta)$ in (\ref{m-eq}) 
in terms of the probabilitiy $W_{ij}^{\rm pq}$ of an inter-site
atomic exchange (``jump'') q$j\leftrightarrow {\rm p}i$ per unit time:
\begin{equation}
W_{ij}^{\rm pq}=n_{{\rm p}i}n_{{\rm q}j}\omega_{ij}^{\rm pq}
\exp[-\beta (E_{{\rm p}i,{\rm q}j}^s- E_{{\rm p}i,{\rm q}j}^{in})]
 \equiv n_{{\rm p}i}n_{{\rm q}j} \gamma_{ij}^{\rm pq}
\exp (\beta  E_{{\rm p}i,{\rm q}j}^{in}).
\label{W_ij}
\end{equation}
Here $\omega_{ij}^{\rm pq}$ and $E_{{\rm p}i,{\rm q}j}^s$
are the ``attempt frequency'' and the ``saddle point energy''
assumed to be independent of alloy configuration;
$\beta =1/T$ is the reciprocal temperature;
$ E_{{\rm p}i,{\rm q}j}^{in}$ is the initial  (before the jump)
configurational energy of jumping atoms, and
$\gamma_{ij}^{\rm pq}=\omega_{ij}^{\rm pq}
\exp[-\beta (E_{{\rm p}i,{\rm q}j}^s)]$ is the
configurationally independent factor in the jump probability.
If we accept for simplicity the pair interaction model, i.e. 
retain only the first  term in Eq. (\ref{H^t}), 
then the operator $ E_{{\rm p}i,{\rm q}j}^{in}$  in (\ref{W_ij})  may be
expressed in terms of formal variational derivatives  of
the hamiltonian (\ref{H^t}) over $n_{{\rm p}i}$ and $n_{{\rm q}j}$, $H_{{\rm p}i}^t=
\delta H^t/\delta n_{{\rm p}i}$ and
$H_{{\rm p}i,{\rm q}j}^t=\delta^2 H^t/\delta n_{{\rm p}i} \delta n_{{\rm q}j}$:
\begin{equation}
E_{{\rm p}i,{\rm q}j}^{in}=
n_{{\rm p}i}H_{{\rm p}i}^t+n_{{\rm q}j}H_{{\rm q}j}^t-
n_{{\rm p}i}n_{{\rm q}j}H_{{\rm p}i,{\rm q}j}^t,
\label{E_ij}
\end{equation}
where the last term corresponds to  the substraction of the ``double-counted''
interaction between the jumping atoms.
The employed neglection of a possible configurational dependence
of $\gamma_{ij}^{\rm pq}$
in (\ref{W_ij}) is actually not essential, and one can  also use
 any form of $W_{ij}^{\rm pq}$ obeying the detailed
balance principle.

It has been shown in \cite{BV-98} that in studies of 
practically interesting problems
 the true vacancy-mediated atomic exchange mechanism
can usually  be replaced by some  equivalent direct exchange model.
For example, instead of a real binary alloy 
ABv with vacancies and the vacancy-mediated
atomic exchanges A$\leftrightarrow$v and  B$\leftrightarrow$v
we can consider a more simple model of a binary alloy AB with the direct 
A$\leftrightarrow$B exchange and only
 one independent variable $n_{{\rm A}i}\equiv n_i$ in Eqs. (\ref{H})--(\ref{E_ij})
for each site $i$. Discussing below for simplicity only the binary alloy  case
we can seek the distribution function $P(\xi)$ in 
(\ref{m-eq}) in the form of a ``generalized Gibbs distribution'':
\begin{equation}
P\{n_i\}=\exp [\beta(\Omega+\sum_{i}\lambda_in_i-Q)].
\label{P}
\end{equation}
Here the operator $Q$ 
is an analogue of the Hamiltonian $H$ in (\ref{H}):
\begin{equation}
Q=\sum_{i>j}a_{ij}n_in_j+\sum_{i>j>k}a_{ijk}n_in_jn_k+\ldots;
\label{Q}
\end{equation}
the ``local chemical potentials'' $\lambda_i$ and
``quasi-interactions'' $a_{i\ldots j}$ 
(being, generally, both time and space dependent)
are the parameters of the distribution; and 
 the generalized grand canonical
potential $\Omega =\Omega\{\lambda_i,g_{i\ldots j}\}$ is determined by the normalizing condition: 
\begin{equation}
\Omega =-T\ln {\rm Tr}\exp [\beta(\sum_{i}\lambda_in_i-Q)].
\label{Omega}
\end{equation}

Multiplying equation (\ref{m-eq}) by operators $n_{i}$, $n_{i}n_{j}$,
etc., and summing over  all configurational states, i.e. over all
number sets $\{n_{i}\}$, we obtain
the set of equations for the averages
$g_{ij\ldots k}=\langle n_in_j\ldots n_k\rangle\equiv {\rm Tr}(n_in_j\ldots n_kP)$, 
in particular, for the mean occupation $c_{i}=\langle n_i\rangle=g_i$.
After certain manipulations described in \cite{Vaks-96,BV-98} these 
equations can be written as:
\begin{equation}
%{d\over dt}
{d\over dt}\,g_{ij\ldots k}=
\sum_s\gamma_{si}\left\langle2\sinh (D_{si}^{-})
 \exp (D_{si}^{+})
n'_{i}n'_{s}n_{j}\ldots n_{k}\right\rangle
+\{i\rightarrow j,\ldots k\}.
\label{g_ik-dot}
\end{equation}
Here $\gamma_{si}$ is $\gamma_{si}^{\rm AB}$; $n'_i$ is $1-n_i$;
$\{ i \rightarrow  j,\ldots k \}$
denotes the sum of expressions obtained
from the first term   by index permutation;
the operators $D_{si}^{ \pm}$ are
\begin{equation}
D_{si}^{ \pm}={1\over 2}\beta[(\lambda_s+(H-Q)_{s}]\pm\{s\rightarrow i\};
\label{D_is}
\end{equation}
and $(H-Q)_{i}$ is the variational derivative $\delta (H-Q)/\delta n_{i}$:
\begin{equation}
(H-Q)_{i}=\sum_{ j>\ldots >k}(v_{ij\ldots k}-a_{ij\ldots k})
n_j\ldots n_k. 
\label{H-Q}
\end{equation}

Eqs. (\ref{P}-\ref{H-Q}) enable us to derive the  microscopic expression for the free energy $F$ of
the nonequilibrium state \cite{Vaks-96,BV-98}:
\begin{equation}
F=\Omega +\sum_i\lambda_i c_i+< H-Q>.
\label{F}
\end{equation}
The function  $F=F\{c_i,g_{i\ldots j}\}$ obeys the generalized first law of thermodynamics,
\begin{equation}%
d F=\sum_i\lambda_idc_i+\sum_{i>\ldots >j}(v_{i\ldots j}-a_{i\ldots j})
dg_{i\ldots j},
\label{dF}
\end{equation}
and has a fundamental property not to 
increase under spontaneous evolution, similarly to the  Boltzmann's not decreasing entropy:
\begin{equation}%
dF/dt\le 0.
\label{dF/dt}
\end{equation}
The stationary state (being not necessarily uniform) corresponds to the minimum of $F$
over its variables $c_i$ and $g_{i\ldots j}$ at the given  $N=\sum_ic_i$:
\begin{equation}%
\partial F /\partial c_i =\lambda_i=\mu; \hspace{ 10mm}
\partial F/\partial g_{i\ldots j}= v_{i\ldots j}-a_{i\ldots j}=0
\label{stationarity}
\end{equation}
where $\mu$ is the Lagrange factor. 
Eqs. (\ref{stationarity}) are the usual Gibbs relations  for the parameters
$\lambda_i$ and $a_{i\ldots j}$ in  the distribution (\ref{P}) for the stationary  case.

%\textheight 220mm
%\newpage

\section{Kinetic mean-field and kinetic cluster approximations}\label{kmfa}

To approximately solve kinetic equations (\ref{g_ik-dot}) one can use
the regular approximate methods of statistical physics, such as the mean-field approximation
(MFA), the cluster variation method (CVM) \cite{Kikuchi,Finel}, and also its  simplified version,
the cluster field method (CFM) \cite{VS}. In  both 
the kinetic MFA and kinetic CFM (KMFA and KCFM) the  equations for 
$c_i(t)$ are separated from those for $g_{i\ldots j}(t)$ and 
have the form \cite{Vaks-96,BV-98,BDPSV}:
\begin{equation}%
dc_i/dt= 2 \sum_jM_{ij}
\sinh [ \beta (F_j-F_i)/2]
\label{c_dot}
\end{equation}
where $F_i=\lambda_i=\partial F/\partial c_i$, while $M_{ij}=M_{ij}\{c_k\}$ and 
$F=F\{c_i\}$ is the MFA or CFM expression for the 
generalised mobility and the free energy  of a nonuniform alloy. 
For simplest approximations, such as the KMFA or the kinetic pair-cluster approximation 
(KPCA), these expressions can be  written analytically. 
 For example, the KMFA 
expressions for  $M_{ij}$ and $F_i$ in an alloy with only pair interactions $V_{ij}^{\rm pq}$
in ({\ref{H^t}) are
\begin{equation}
M_{ij}=\gamma_{ij}\sqrt{c_ic_i'c_jc_j'}
\exp \left[{\beta\over 2} \sum_k(u_{ik}+u_{jk})c_k\right];\quad 
F_i=\ln {c_i\over c_i'}+\sum_kv_{ik}c_k.
\label{KMFA}
\end{equation}%
Here $c_i'=1-c_i$, and $u_{ik}=V_{ik}^{\rm AA}-V_{ik}^{\rm BB}$ is the ``asymmetrical
potential'' \cite{Martin}. 
Substituting Eqs. (\ref{KMFA})
into (\ref{c_dot}) we obtain the analytical KMFA equation for $c_i(t)$.
A similar equation is obtained in the KPCA \cite{BV-98}.  

The usual phenomenological
kinetic equations, in particular, those used by Khachaturyan and coworkers
\cite{Khach-book,CWK-92,Khach-00}, correspond to 
the  linearization of the KCFM equations (\ref{c_dot}) in $(F_i-F_j)$ and neglecting the 
$c_i$-dependence in the mobility $M_{ij}$. Such approach is usually
sufficient for qualitative considerations, but for some problems it can lead to a notable 
distortion of both the time scale and other details of the microstructural evolution
\cite{DVM}. 

 KMFA or KCPA  are usually sufficient for studies of main kinetic features of
 spinodal decomposition, as well as orderings in the BCC lattice
 \cite{DMSV,DVM,BV-96,BV-97,DV,BSV}.
 However, for more complex orderings, e.g. L1$_2$ and  L1$_0$ orderings  
in the FCC  lattice, MFA and PCA are known to be insufficient and 
more precise methods are necessary, 
such as the CVM or CFM \cite{Kikuchi,Finel,VS}. 
Recently we suggested a simplified version of CVM, the tetrahedron cluster
field method (TCFM) \cite{VS},
that combines a high accuracy  of CVM in describing thermodynamics 
 with great simplification of calculations making it posible
to develop  its kinetic generalization, KTCFM \cite{BDPSV}. 
Similarly to the KMFA and KPCA, the KTCFM provides explicit equations for
 the mobility $M_{ij}$ and the local chemical potential $F_i$ in
Eq. (\ref{c_dot}) via mean occupations $c_k$, and for each site $i$ these 
equations can be reduced to  a system  of four nonlinear algebraic equations 
which can easily be solved  using  Newton's method.

Let us now make remarks about effective interactions
$v_{i\ldots j}$ in the Hamiltonian  (\ref{H}) for real alloys.
These interactions include the
``chemical'' contributions $v_{i\ldots j}^c$
which describe the energy changes under permutations of  atoms A and B
in the rigid lattice, and the 
 ``deformational'' interactions $v_{i\ldots j}^d$
related to the  lattice deformations under such permutations.
The chemical contributions are estimated from either first-principle
calculations or fitting to some experimental data \cite{Finel}, but  
for the long-ranged deformational interactions 
such methods can not be directly used.
 A microscopical  model for $v^d$ in dilute alloys which 
includes only one experimental parameter
 was suggested by Khachaturyan \cite{Khach-book}. 
The deformational interaction in concentrated alloys can lead to some new 
effects being absent in dilute alloys, for example, to the
lattice symmetry changes under phase transformations, such as 
the tetragonal distortion under L1$_0$ ordering, and earlier these effects
 were described only phenomenologically \cite{CWK-92}.
Recently we suggested  a microscopical model for calculations of $v^d$
in concentrated alloys \cite{BDSV} which generalizes
 the Khachaturyan's approach \cite{Khach-book}.
Unlike the case of dilute alloys, this deformational interaction 
turns out to be  essentially non-parwise, and it
includes two parameters which can be found from experimental data
about the lattice distortion under phase transformations.

\section{Methods of simulation of L1$_0$ ordering}\label{methods}

Up to recently most of theoretical treatments of kinetics of alloy ordering 
considered only simplest B2 (CuZn-type) orderings with just two types of 
antiphase-ordered domain (APD) and one type of 
antiphase boundary (APB) separating these APDs.
 Yet  ordered structures in real alloys are usually
more complex and include many types of APD.
For example, under the D0$_3$ (Fe$_3$Al-type) 
ordering on the BCC lattice there are four types of APD \cite{BSV},
while under the L1$_2$ (Cu$_3$Au-type) or L1$_0$ ordering on the FCC lattice there are
four or six types of APD, respectively.
It  results in a number of peculiar kinetic
features that are absent for the simple B2 ordering. In Refs.~\onlinecite{BSV,BDPSV}
we discussed such features for the D0$_3$ and 
L1$_2$-type orderings.
Below we  consider the L1$_0$-type orderings for  which the
microstructural evolution  turns out to be still more complex and interesting.

To study this evolution we made
simulations of  A1$\to$L1$_0$
 transformations after a quench of
an alloy from the disordered 
A1 phase to the single-phase L1$_0$ state. 
For these simulations we used  the master equation approach and
the KTCFM described above. We considered  five  alloy models
with different types of chemical interaction: 
the second-neighbor 
(or ``short-range'') interaction models 1, 2, and 3 with $v_1=1000$ K and the ratio
$v_2/v_1$ equal to (-0.125), (-0.25) and (-0.5), respectively;
the ``intermediate-range'' fourth-neighbor-interaction model 4 
with $v_n$ estimated by Chassagne et al. 
\cite{Chassagne} from their experimental data for Ni--Al alloys:
 $v_1=1680\,\mathrm{K}$, $v_2=-210\,\mathrm{K}$,
$v_3=35\,\mathrm{K}$, and $v_4=-207\,\mathrm{K}$;
 and the ``extended-range'' fourth-neighbor-interaction model 5
with $v_1=1000$ K, $v_2/v_1=-0.5$,   $v_3/v_1=0.25$,
 and  $v_4/v_1=-0.125$. 
The deformational interaction  $v^d$ for all these models
was estimated as described in section 3
with the use of experimental data for Co-Pt alloys.
The critical temperature  $T_c$ for our models
 corresponds
to the stoichiometric composition $c=0.5$ and 
is 614, 840, 1290, 1950 and 2280 K for model 1, 2, 3, 4 and 5, respectively.

The distribution of mean occupations $c_i$ under alloy ordering can be
described in terms of both long-ranged and local order parameters. 
For the homogeneous L1$_2$ or
L1$_0$ ordering the distribution $c_i=c\hspace{1pt} ({\bf R}_i)$ (where ${\bf R}_i$ 
is the FCC lattice vector) can be written 
in terms of three long-ranged order parameters $\eta_{\alpha}$,
see \hbox{e. g. \cite{Khach-book}}:
\begin{equation}
c_i=c+\eta_1\exp (i{\bf k}_1{\bf R}_i)+\eta_2\exp (i{\bf k}_2{\bf R}_i)
+\eta_3\exp (i{\bf k}_3{\bf R}_i)
\label{c_ordered}
\end{equation}
where $\bf k_{\alpha}$ is the superstructure vector corresponding
to $\eta_{\alpha}$: \ \ $\{{\bf k_1, k_2, k_3}\}=\{[100],[010],[001]\}2\pi/a$.
For the cubic L1$_2$  structure $|\eta_1|=|\eta_2|=|\eta_3|$, $\eta_1\eta_2\eta_3>0$,
and four types of ordered domain are possible. In the L1$_0$-ordered structure 
with the tetragonal axis $\alpha$, a single non-zero 
parameter $\eta_{\alpha}$ is present which is either positive or negative.
Therefore, six types of ordered domains are possible with two types of APB.
That separation of two APDs with the same  tetragonal axis will 
be for brevity called the ``shift-APB'', and that separation of the APDs with 
perpendicular tetragonal axes will be called the ``flip-APB''.
The transient partially ordered alloy states 
can be conveniently described in terms of 
local squared order parameters  $\eta_{\alpha i}^2$  defined in \cite{BDPSV}.
The simulation results in figures 1--5 below are usually presented as the
distributions of quantities $\eta_i^2=\eta_{1i}^2+\eta_{2i}^2+\eta_{3i}^2$
(to be called the ``$\eta^2$--representation''):
the grey level linearly varies with $\eta_i^2$ between its minimum and
maximum values from completely dark to completely bright, and this
distribution is similar to that observed in the transmission electron
microscopy (TEM) images \cite{Tanaka,Zhang,Yanar,Leroux,Guymont}.

The simulations were performed in FCC simulation boxes of sizes
$V_b=L^2\times H$ (where $L$ and $H$ are given in units of the lattice constant $a$)
with periodic boundary conditions.
We used both 3D simulations with $H=L$ and quasi-2D simulations
with $H=1$, and all significant features of evolution 
in both types of simulation were found to be similar.
Each of figures 1--5 below includes all FCC lattice 
sites lying in two adjacent planes, $z=0$ and $z=a/2$, 
thus it shows $4L^2$  lattice sites.
The initial as-quenched
distribution $c_i(0)$ was characterized by its mean value $c$ and small
random fluctuations $\delta c_i$; usually we used $\delta c_i=\pm 0.01$.

\section{Kinetics of  A1$\to$L1$_0$ transformation}

 To avoid discussing the problems of nucleation, in this work we consider
the transformation temperatures $T$ 
lower than the ordering spinodal temperature $T_s$.
Then the  evolution under the A1$\to$L1$_0$  transition includes
the following stages  \cite{Tanaka,Zhang,Yanar,Leroux}:

(i) The initial stage of the formation of  finest  L1$_0$-ordered domains
when  their tetragonal distortion  still has  little effect on the evolution
and all six types of APD are present in microstructures in the same proportion.

(ii) The  imtermediate stage which corresponds to the so-called ``tweed'' contrast in 
 TEM images. The tetragonal deformation of the
L1$_0$ phase here  leads to a predominance of the (110)-oriented 
flip-APBs in the microstructures, but all six types of APD 
 are still present   in  similar proportions.

(iii) The final, ``twin'' stage when the tetragonal distortion
of the L1$_0$-ordered APDs  becomes
the main factor determining the evolution and leads to the
formation of the (110)-type oriented bands.
 Each band includes only two types of APD with the same tetragonal
axis, and  these axes in the adjacent bands  are ``twin'' related, i.e.  have the 
alternate (100) and (010) orientations for the given set of the (110)-oriented bands.

The thermodynamic driving force for the (100)-type orientation of flip-APBs
is the gain in the elastic energy of the adjacent APDs: at other orientations
 this  energy  increases under the growth of an APD
proportionally to its volume \cite{Khach-book,Vaks-01}.
For an APD with the characteristic size $l$, surface $S_d$, tetragonal deformation
$\varepsilon$, and shear constant $c_s$,  this force
begins to affect the microstructural  evolution when the volume 
elastic energy $E_{el}^v\sim
c_s\varepsilon^2lS_d$ becomes comparable with the surface energy
$E_s\sim\sigma S_d$ where $\sigma$ is the APB surface energy.
 The beginning of the tweed stage (ii) corresponds to the relation
$E_{el}^v\sim E_s$ or to the characteristic 
 APD size  
\begin{equation}
l_0\sim \sigma/c_s\varepsilon^2.
\label{l_0}
\end{equation}
The distortion $\varepsilon$ is proportional to the 
order parameter squared  \cite{Vaks-01}, 
and below it is  characterized by its maximum  value 
$\varepsilon_m$ corresponding to $T=0$ and $c=0.5$.

Some results of our simulations are presented in figures
\ref{m5s_e10}--\ref{m2_e10}. The symbol A or
$\overline {\rm A}$, B or $\overline {\rm B}$ and C or $\overline {\rm C}$
in these figures corresponds to
an L1$_0$-ordered domain with the tetragonal axis along (100), (010) and (001)
and the positive or negative value of  the order parameter  $\eta_{1}$, $\eta_{2}$ 
and $\eta_{3}$, respectively. Frame \ref{m5s_e15}d is shown in 
the $\eta_{2}^2$-representation:  the grey level linearly varies with 
$\eta_{2i}^2$ between its minimum and
maximum from completely dark to completely bright,
which corresponds to the usual bright-field TEM images 
\cite{Tanaka,Zhang,Yanar,Leroux,Guymont}.

\begin{figure}
\begin{center}
\epsfig{file=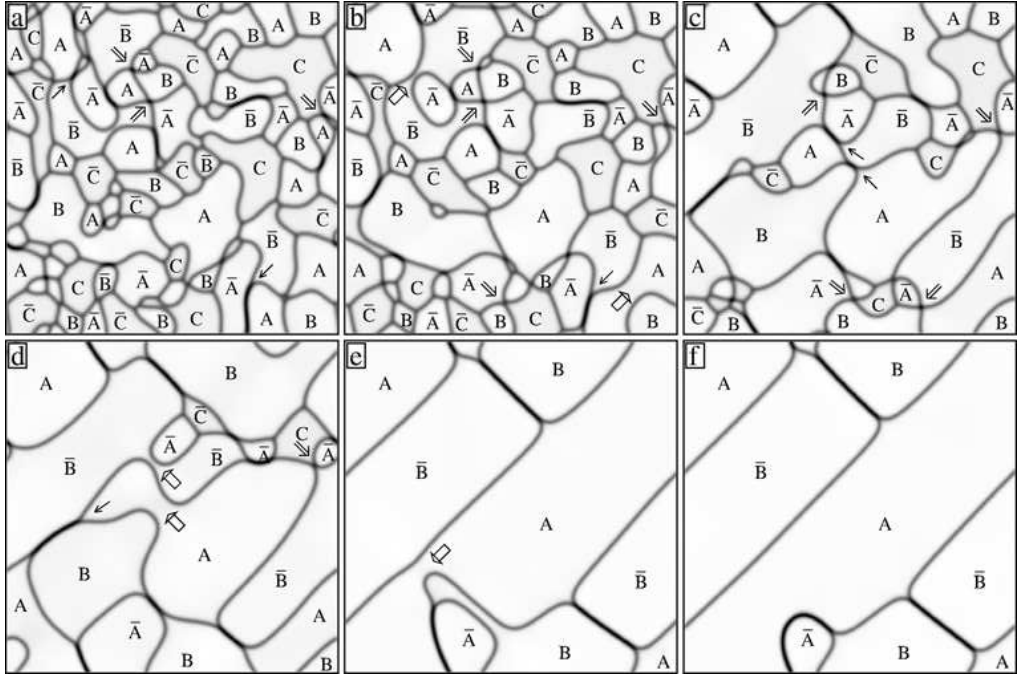,angle=-90,width=0.75\textwidth}
\end{center}
\caption{Temporal evolution of the extended--range-interaction model~5
under the phase transformation A1$\to$L1$_0$ shown in the $\eta^2$-representation
for the simulation box size $V_b=128^2\times 1$ at 
$c=0.5$, the reduced temperature $T'=T/T_c=0.7$,  $|\varepsilon_m | =0.1$
and the following values of the reduced time $t'=t\gamma_{ij}$: (a)~10; (b)~20;
(c)~50; (d)~100; (e)~250; and (f)~280. The symbol A,
$\overline {{\rm A}}$, B, $\overline {\rm B}$, C or $\overline {\rm C}$ indicates 
the type of the ordered domain as described in the text.
The single, the double and the  thick arrow indicates  the splitting APB process, 
the quadruple junction of APDs, and the fusion-of-domain process,
respectively, mentioned in the text.}
\label{m5s_e10}
\end{figure}

\begin{figure}
\begin{center}
\epsfig{file=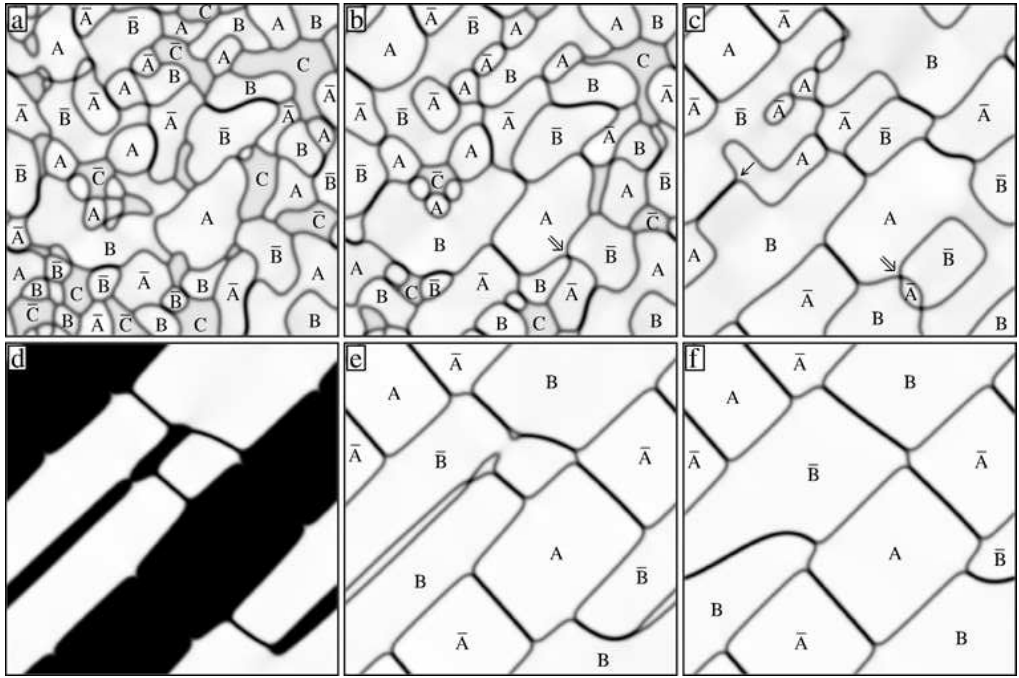,angle=-90,width=0.75\textwidth}
\end{center}
\caption{As  figure \ref{m5s_e10}, but
at $|\varepsilon_m | =0.15$
and the following values of  $t'$: (a)~10; (b) 20; (c)~50; (d)~150; (e)~172;  and (f)~350.
Frame 2d is shown in the $\eta_2^2$-representation described in the text.
\label{m5s_e15}}
\end{figure}

Figures 1--5 illustrate quasi-2D simulations for which microstructures include
only the edge-on APBs normal to the (001) plane. The 
above-mentioned  elimination of the 
volume-dependent elastic energy in such geometry
is possible only for the (100) and (010)-ordered APDs
A or $\overline {\rm A}$ and B or $\overline {\rm B}$
separated by the (110)-oriented flip-APB, while in 
the (001)-oriented domains C and
$\overline {\rm C}$ this elastic energy is always present. Therefore, 
the tweed stage (ii) in these simulations  corresponds to both the
predominance of  (110)-oriented APBs
and the decrease of the number of domains C and $\overline {\rm C}$  in 
the microstructures.

\begin{figure}
\begin{center}
\psfig{file=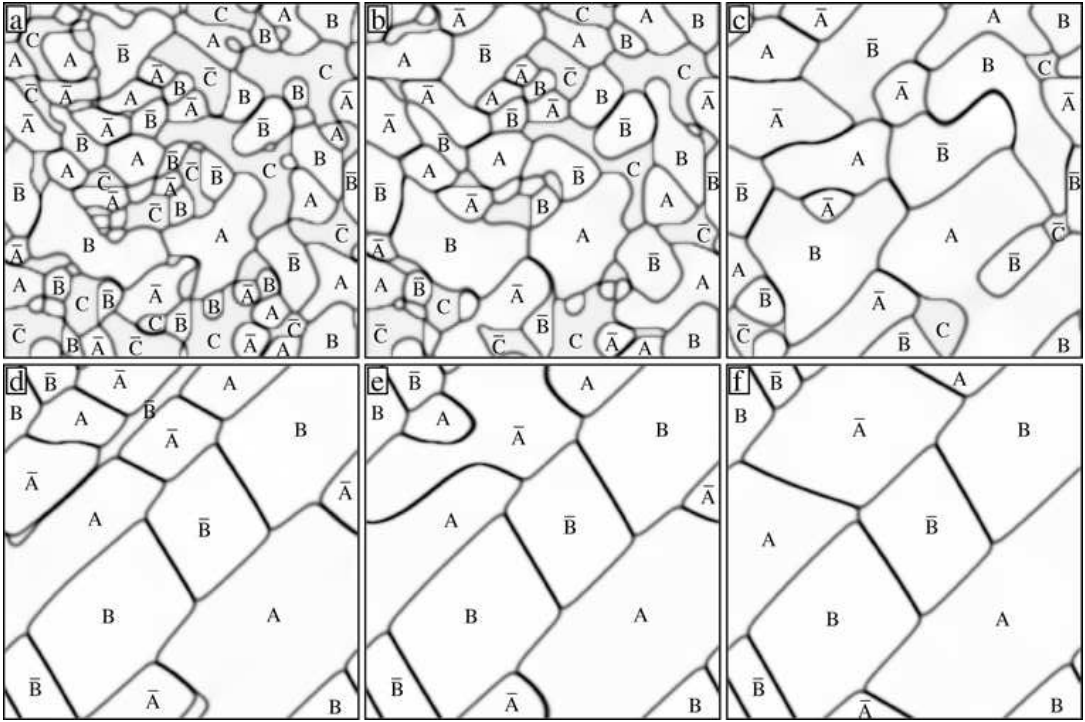,angle=-90,width=0.8\textwidth}
\end{center}
\caption{
As  figure \ref{m5s_e15}, but 
for the intermediate-range-interaction model 4 at $T'=0.67$ 
and the following values of $t'$: 
(a)~10; (b)~20; (c)~50; (d)~170; (e)~200;  and (f)~700.
\label{m4s_e15}}
\end{figure}

Let us first consider  figures \ref{m5s_e10}--\ref{m4s_e15} in which
frames \ref{m5s_e10}a--\ref{m5s_e10}b, \ref{m5s_e15}a and \ref{m4s_e15}a
correspond to the initial stage; frames \ref{m5s_e10}c--\ref{m5s_e10}d,
 \ref{m5s_e15}b--\ref{m5s_e15}c, and \ref{m4s_e15}b--\ref{m4s_e15}c,
to the tweed stage; and the rest frames, to the twin
stage. At both the initial  and the tweed stage we can observe
the following  features of evolution  \cite{PV}:

(a) The presence of abundant processes of fusion of in-phase domains which are
 one of the main mechanisms of domain growth  at these stages.

(b) The presence of  peculiar long-living configurations, the
 quadruple junctions of APDs (4-junctions) of the type
A$_1$A$_2$$\overline{\rm A_1}$A$_3$ where A$_2$ and A$_3$ can correspond
to any two of four types of APD different from A$_1$ and $\overline{\rm A_1}$.

(c) The presence of many processes of ``splitting'' of a shift-APB into two 
flip-APBs which is finished by either a fusion of in-phase domains 
mentioned in point (a) ($s\to f$ process), or a formation of a 4-junction 
mentioned in point (b) ($s\to 4j$ process).

For example, $s\to f$ processes
can be followed in  frames 
\ref{m5s_e10}a--\ref{m5s_e10}b; \ref{m5s_e10}c--\ref{m5s_e10}d;
\ref{m5s_e10}d--\ref{m5s_e10}e; \ref{m5s_e15}c--\ref{m5s_e15}d; etc. 
The fusion with the disappearance
of an intermediate APD which initially separates two in-phase domains to be fused \cite{PV}
 can be seen in the lower right part of frames \ref{m5s_e10}a--\ref{m5s_e10}b.
Several long-living 4-junctions  are seen in frames 
\ref{m5s_e10}a--\ref{m5s_e10}d and \ref{m5s_e15}c--\ref{m5s_e15}d;
 and an $s\to 4j$ process can be followed
in the lower right part of frames \ref{m5s_e10}a--\ref{m5s_e10}c.
Let us also note that the microstructural features (b) and (c)  can
be naturally explained by a significant excess of the surface energy of shift-APBs 
with respect to flip-APBs
found in our CFM calculations for the systems under consideration.

Frames \ref{m4s_e15}a--\ref{m4s_e15}c also display 
some (100)-oriented and thin ``conservative'' APBs \cite{BDPSV,PV}. Such   APBs 
are most typical of the short-range-interaction systems---see \cite{BDPSV} and
figure \ref{m1s_e15} below---where they have a low surface energy 
(being zero  for  the nearest-neighbor interaction model).
Under an increase of the interaction range (as well as temperature or the deviation 
from stoichiometric composition $\delta c=0.5-c$) the anisotropy of this surface energy 
decreases,  and so for  model 4 
such APBs are few,
while for the extended-range-interaction model 5 they are absent at all.

Frames \ref{m5s_e10}c--\ref{m5s_e10}d,  \ref{m5s_e15}b--\ref{m5s_e15}c,
and \ref{m4s_e15}b--\ref{m4s_e15}c (as well as \ref{m1s_e15}a--\ref{m1s_e15}b)
illustrate a (110)-type alignment of APBs between APDs A or $\overline {\rm A}$ 
and B or $\overline {\rm B}$ and a ``dying out'' of APDs C and $\overline{\rm C}$
  at the tweed stage. Frames \ref{m5s_e10}c and \ref{m5s_e10}d
also show  that in the simulation with a realistic distortion parameter
 $|\varepsilon_m|=0.1$ (fitted to the structural data for CoPt)
the  APD size $l_0$ (\ref{l_0}) characteristic 
of the tweed stage is about $(20-40)\,a$. It agrees with the order of magnitude 
of $l_0$ observed in the CoPt-type alloys FePt and FePd \cite{Zhang,Yanar}.

Discussing the final, twin stage of the evolution we first mention some characteristic 
configurations observed in experimental studies of transient twinned microstructures
\cite{CWK-92,Zhang,Yanar,Leroux,Guymont}:

(1) ``semi-loop'' shift-APBs adjacent to the twin band boundaries;

(2) ``S-shaped'' shift-APBs stretching across the twin band;

(3) short  and narrow twin bands---usually with one or two 
shift-APBs near their edges--lying within the main twin bands;

(4) an alignment of shift-APBs in the final,  ``nearly equilibrium'' twin bands: 
within a (100) oriented  band in a (110)-type
polytwin the APBs tend to align normally to 
some direction ${\bf n}=(\cos\alpha, \sin\alpha, 0)$ with a  ``tilting''
 angle $\alpha$ which is less than 
$\pi/4$ in CoPt  and is close to zero
in CuAu \cite{Leroux,Guymont}. 

Comparing our results with experiments
one should consider that  due to the limited size of the simulation box
the twin band width $d$ in our simulations
 has the same order of magnitude as the 
APD size $l_0$ (\ref{l_0}) characteristic of the tweed stage, while 
in experiments $d$ usually much exceeds $l_0$ \cite{Zhang,Yanar,Leroux,Guymont}.
Therefore, the distribution of shift-APBs within twin bands 
in our simulation is usually much more close to  equilibrium than in experiments.
In spite of this difference,  the simulations reproduce not only 
 ``nearly equilibrium'' configurations (4) 
but also transient configurations (1)--(3)  and elucidate their formation mechanisms.
In particular, both the ``semi-loop'' and ``S-shaped'' shift-APBs
are  formed from regular-shaped approximately quadrangular APDs
(characteristic of early twin stages) due to the disappearance of
adjacent APDs which are ``wrongly-oriented'' with respect to the given twin band;
it is seen, for example, in frames  \ref{m5s_e10}d--\ref{m5s_e10}f,
 \ref{m5s_e15}e--\ref{m5s_e15}f,  \ref{m4s_e15}c--\ref{m4s_e15}e, etc.
The formation of short and narrow twin bands with the edges touched by
shift-APBs is illustrated by frame \ref{m5s_e15}d (which is strikingly similar
to the experimental microstructure shown in Fig.~2 of Ref.~\onlinecite{CWK-92}) and also by frame
\ref{m4s_e15}d. 

The alignment of shift-APBs mentioned in point (4)
is illustrated by frames \ref{m5s_e15}f and \ref{m4s_e15}f, as well as 
 \ref{m1s_e15}d and  \ref{m2_e10}a--\ref{m2_e10}d.  These frames
show that the tilting angle $\alpha$ sharply depends on the interaction type,
particularly on the interaction range, as well as on the concentration $c$ and temperature $T$.
In particular, for the extended-range-interaction model 5 this angle is close 
to  $\pi/4$; for the intermediate-range-interaction model 4 (which seems to more
realistically describe properties of CoPt-type alloys) the angle 
$\alpha$ is less than $\pi/4$, i.e. APB planes are tilted 
to the tetragonality axis; and  for the short-range-interaction models 1 and 2 
the APB  planes tend to be parallel with the tetragonality axis, i. e. $\alpha\simeq 0$.
Such interaction-dependent alignment of shift-APBs  can be
explained \cite{Vaks-01} by the competition between the anisotropy of their 
surface energy 
--- which for both the intermediate and short-ranged-interaction
systems tends to orient APBs parallel with the tetragonality axis decreasing
the angle $\alpha$, and a tendency to minimize the APB area within the given twin band which
corresponds to $\alpha =\pi/4$. Therefore, the comparison of
experimental tilting angles with the theoretical calculations \cite{Vaks-01}
can provide both qualitative and quantitative
information about the interatomic interactions in an alloy.

\begin{figure}
\begin{center}
\psfig{file=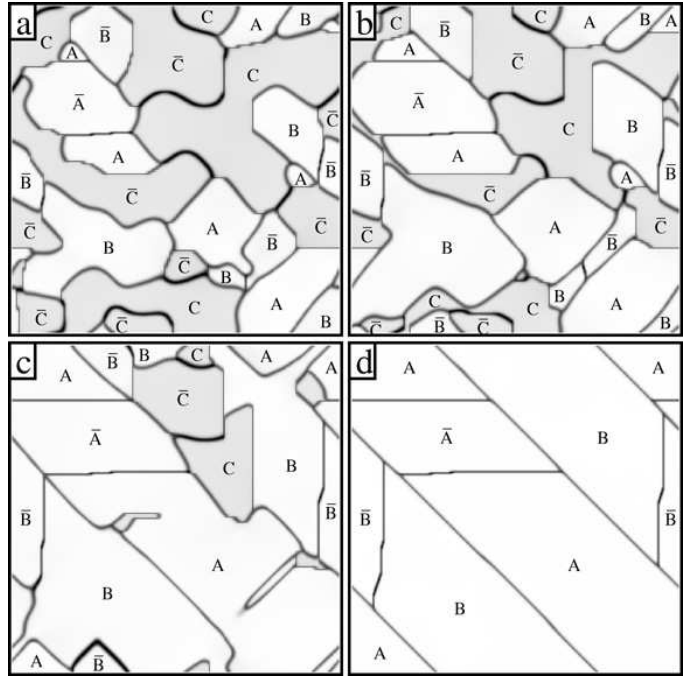,angle=0,width=0.5\textwidth}
\end{center}
\caption{As figure \ref{m5s_e15} but for the short-range-interaction model 1
at $T'=0.9$ and the following values of $t'$: (a) 30;  (b) 40; (c) 60; and (d) 120.
\label{m1s_e15}}
\end{figure}
 
Figure  \ref{m1s_e15} illustrates evolution for model 1
which describes the short-ranged-interaction  systems such as alloys
 Cu--Au  \cite{BDPSV}. The microstructures for such systems include 
many conservative APBs mentioned above,
and the shift-APBs in the final frame \ref{m1s_e15}d are ``step-like'' 
consisting of (100)-type oriented conservative segments and small
non-conservative ledges (being similar to the APBs observed  in the
 L1$_2$-ordered Cu$_3$Au alloy \cite{BDPSV}). These step-like APBs 
can be viewed as  a ``facetted'' version of  tilted APBs seen in 
frames \ref{m5s_e15}f,  \ref{m4s_e15}f,  \ref{m2_e10}c and \ref{m2_e10}d. 
As mentioned, under an increase of temperature $T$ or 
 ``non-stoichiometry'' $\delta c= 0.5-c$ 
 the anisotropy in the APB energy rapidly decreases.
 It results in  sharp, phase-transition-like
changes in morphology of aligned APBs,  from  the  ``faceting'' to the
``tilting'', which is illustrated by frames \ref{m2_e10}a--\ref{m2_e10}d. These
morphological changes are  realized via some local bends 
of facetted APBs illustrated by frames  \ref{m2_e10}a--\ref{m2_e10}b.
Therefore, this ``morphological phase transition'' is 
actually smeared over some intervals of temperature or concentration,
but frames  \ref{m2_e10}a--\ref{m2_e10}d show that these intervals  can be
relatively narrow.

\begin{figure}
\begin{center}
\psfig{file=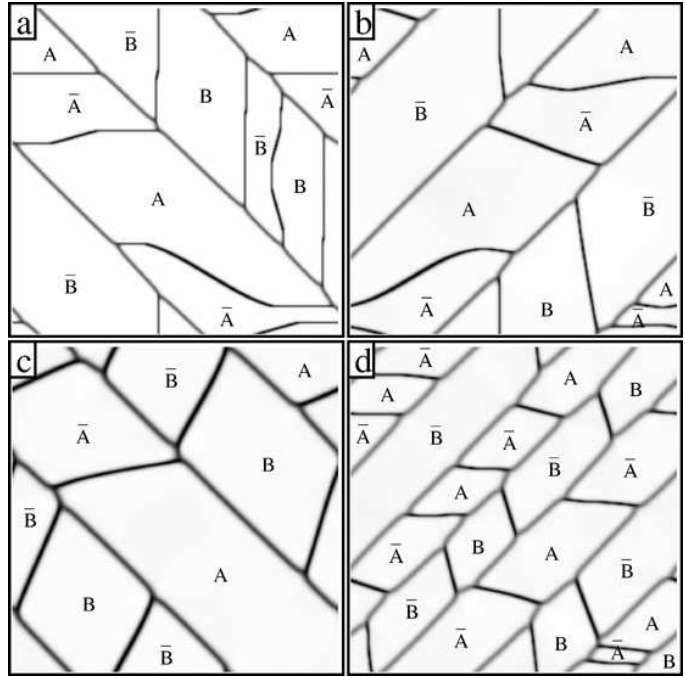,angle=0,width=0.5\textwidth}
\end{center}
\caption{As figure \ref{m5s_e10} but for model $2$
at  $|\varepsilon_m|=0.1$  and the following values of  $c$, $T'$ and $t'$: 
(a) $c=0.5$, $T'=0.77$, $t'=350$;
(b) $c=0.46$, $T'=0.77$, $t'=350$;
(c) $c=0.5$, $T'=0.95$, $t'=300$;
and (d) $c=0.44$, $T'=0.77$, $t'=300$.
\label{m2_e10}}
\end{figure}

Finally, let us make a general remark about the kinetics 
of multivariant orderings in alloys, such as the D0$_3$, L1$_2$ and
L1$_0$ orderings considered in Refs.~\onlinecite{BDPSV,BSV,Vaks-01,PV}
and in this work.
It is well known that the thermodynamic behavior of different systems
under various phase transitions reveals features of universality and 
insensitivity to microscopical details of structure,
 particularly in the critical region near thermodynamic instability points. 
The results of this and other studies of  multivariant oderings  
show that such universality 
does not seem to hold for their phase transformation kinetics,
 anyway outside the critical region (which for 
such orderings  is usually
either quite narrow or absent at all). The microstructural evolution 
 reveals a great variety of peculiar features,
the detailed form of which sharply depends on the type of 
interatomic  interaction, structure, the degree of deviation from stoichiometric composition,
  and other ``non-universal'' characteristics.

\section*{Acknowledgments}

The authors are much indebted to Georges Martin for numerous stimulating discussions.
The work was supported  by the Russian Fund of Basic Research under
Grants No. 00-02-17692 and 00-15-96709.

\end{document}